\documentstyle[prl,aps,twocolumn]{revtex}
\begin{document}

\twocolumn[\hsize\textwidth\columnwidth\hsize\csname@twocolumnfalse\endcsname
\title{{\large {\bf Jastrow-Luttinger Fractional Liquids}}}
\author{K.-V. Pham, M. Gabay and P. Lederer}
\address{Laboratoire de Physique des Solides, associ\'e au CNRS \\
Universit\'{e} Paris--Sud \\
91405 Orsay, France}
\date{\today}
\maketitle
\begin{abstract}
{
In this paper, we present a description of Haldane's Luttinger liquid which
parallels Laughlin's theory of the Fractional Quantum Hall (FQH) incompressible
fluid, both exhibiting similar ground states as well as fractional
excitations. These two non-Fermi liquids are instances of a generic
structure for low-dimensional quantum liquids which we propose to dub 
Jastrow-Luttinger
Fractional Liquids. An important feature of such liquids is the complete
 fractionalization of the parent particle. In particular, in both one and {\it
two} dimensions spin-charge separation can be achieved and is indeed suggested 
to
occur for unpolarized quantum Hall systems both at the edge and in the bulk.
}
\end{abstract}
\pacs{PACS Numbers: }
] \widetext
\narrowtext
The goal of this paper is to present strong arguments supporting the
existence of a generic theoretical structure for a class of low dimensional
condensed matter systems, which we propose to dub Jastrow-Luttinger
fractional liquids. Among these systems we find two prominent non-Fermi
liquids namely the Luttinger liquid (LL) which describes a one dimensional
metal\cite{haldane}, and in two dimensions the incompressible fluid of the
Fractional Quantum Hall Effect (FQHE)\cite{laughlin}. The fact that these
two quantum liquids are related is well known: indeed edge excitations of a
quantum Hall sample are believed to be described by a chiral Luttinger liquid%
\cite{wen}, a variant of the usual Luttinger liquid. Yet, it will perhaps
come as a surprise to learn that the connection between the two fluids is
more fundamental. Indeed the wavefunctions of both the ground state and
excitations of the gaussian boson hamiltonian describing the LL are the
precise 1D analogs of the variational states considered in the FQHE: the
ground state is a Jastrow-Laughlin (JL) wavefunction, charged excitations
have the functional form of Laughlin's quasiparticles, while neutral
excitations are density fluctuations in complete analogy to the single-mode
approximation (SMA) approach to the FQHE \cite{plat,prange}. Our claim is
consistent with the result obtained by Fradkin et al \cite{fradkin} who used
a general field theoretical formalism to show that the square modulus of the
Thirring model ground state functional has the Jastrow-Laughlin form. We
will actually see that the full wavefunctions of both the ground state and
its excitations can be derived in a very elementary manner.

We use the term ''fractional'' to describe Jastrow-Luttinger liquids because
they exhibit a {\it fractionalization} of the basic constituent particles:
the quantum numbers of these particles (electrons, bosons or spins) have
completely vanished from the spectrum of elementary excitations\cite
{anderson97}. This is first evidenced at the level of charged excitations
which no longer carry a unit charge with respect to the ground state (e.g.
the fractional charges of the FQHE, the spinon of the 1D Heisenberg model).
Furthermore when we add internal quantum numbers to the $U(1)$ charge, the
hamiltonian and neutral excitations show a separation into independent
collective modes each separately carrying part of the quantum numbers of the
basic particle; this separation for the hamiltonian carries over to the
ground state and charged excitations wavefunctions, and to all quantum
averages: they can be factorized into dynamically independent parts. We
usually have independent charge and spin modes -that is spin-charge
separation as in the LL- if some simple symmetry requirement (derived below)
is obeyed; if the condition is not fulfilled, spin-charge separation is not
realized but a more general {\it quantum numbers separation} still exists.
An instance will be given with the one dimensional Hubbard model in a
magnetic field. After a discussion of the 1D case with the LL, we will turn
to the 2D case. We will show that spin-charge separation may occur in
quantum Hall samples whenever a spin unpolarized ground state is achieved,
for states describable by a Halperin wavefunction (a multicomponent
generalization of the usual Laughlin state\cite{halperin}): this is
experimentally relevant to fillings such as $\nu =8/5$. We will also define
the idea of {\it pseudo-confinement }which is required for a correct
understanding of fractionalization. After a discussion of edge states in the
FQHE whose microscopic relation to Laughlin's bulk theory we try to clarify,
we briefly discuss issues in quantum magnetism such as the spin liquid,
arguing for its existence. {\bf {\centerline {Heisenberg Model.}}} Our
initial clue came from the observation that both the one dimensional
Heisenberg chain and the $XY$ model - which are Luttinger Liquids - have
ground states (approximate in the first case) with the Jastrow-Laughlin (JL)
form: ${\psi }_{\lambda }^{{}}(\{r_{1,..,}r_{N}\})=\prod_{i<j}\mid z_{ij}{%
\mid }^{\lambda }$ ,where the $\{r_{i}\}$ are the positions of $N$ down
spins (in the hard-core boson representation) on a chain of length $L=2N$
with periodic boundary conditions, and $z_{i}=e^{i2\pi r_{i}/L}$ ($\lambda =1
$ for the $XY$ model, while $\lambda =2$ for the Heisenberg chain). The
approximate ground state for the Heisenberg chain yields a very good
variational energy\cite{Horsch83} and its correlators have the correct large
distance properties: it is actually the exact ground state of a $1/r^{2}$
exchange spin model, the Haldane-Shastry chain which belongs to the
universality class of the Heisenberg model\cite{hal88}. The
Calogero-Sutherland model\cite{cs}, a $1/r^{2}$ interaction model related to
the Haldane-Shastry chain was shown through finite-size scaling methods
combined to conformal theory to be a Luttinger liquid\cite{kawakami}: again
its exact ground state has the Jastrow-Laughlin form. These remarks led us
to investigate whether Jastrow-Laughlin wavefunctions might be suitable
variational states for arbitrary Luttinger liquids. We tested this idea on
the anisotropic Heisenberg chain: 
\begin{eqnarray}
{H}_{xxz}(\Delta ) &=&J\sum_{i}\frac{1}{2}\ (S_{i}^{+}\
S_{i+1}^{-}+S_{i}^{-}\ S_{i+1}^{+})+\Delta \ S_{i}^{z}\ S_{i+1}^{z} 
\nonumber \\
&=&J\sum_{i}\frac{-1}{2}\ (b_{i}^{+}\ b_{i+1}^{{}}+b_{i}^{{}}\
b_{i+1}^{+})+\Delta \ n_{i}\ n_{i+1}  \label{xxz}
\end{eqnarray}
where in the second line we use a hard-core boson representation for $S=1/2$
spins \cite{auerbach}. (A rotation around the $z$ axis has also been
performed for spins on odd sites.) $J$ is the exchange integral and the
anisotropy $\Delta $ is chosen to vary in the range $[-1;1]$ where the above
model is known to be a LL\cite{luther}. We then consider : 
\begin{equation}
{\psi }_{\lambda }^{{}}(\{r_{1,..,}r_{N}\})=\prod_{i<j}\mid z_{ij}{\mid }%
^{\lambda }  \label{three}
\end{equation}
and find indeed that $\Psi _{\lambda }$ yields excellent variational
energies per site correct to the third decimal place when compared to the
exact ground state energies obtained from the Bethe ansatz solution. It is
known that the Luttinger liquid parameter $K$ which controls the anomalous
exponents is given for a given anisotropy $\Delta $ by $\Delta =-\cos (\frac{%
\pi }{2K})$\cite{luther}. We find that the best values of $\lambda $ for a
given $\Delta $ satisfy $\lambda =1/K$. This is expected since a 1D plasma
analogy similar to the one introduced in\cite{laughlin} gives that the
transverse spin-spin correlator for $\Psi _{\lambda }$ decays as $%
1/r^{\lambda /2}$ at large distance while the LL theory predicts an exponent 
$1/2K$ \cite{schulz}. (In one dimension the fictitious plasma is the
well-known Dyson gas of random matrices\cite{dyson}.) {\bf {\centerline {%
Ground state and neutral excitations of the LL.}}} The previous result
suggests that all Luttinger liquids might indeed be liable to a variational
Jastrow-Laughlin description, which leads us to address the question of the
precise relation beween such a variational approach and the LL formalism. In
other words we would like to compare our Jastrow wavefunction with the
ground state of the boson LL hamiltonian. The task is easy; the gaussian
boson hamiltonian is just a sum of harmonic oscillators so that the
determination of the ground state and of its excitations is trivial: 
\begin{equation}
H_{B}=\frac{u}{2}\int_{0}^{L}dx\ K^{-1}(\nabla \Phi )^{2}+K(\nabla \Theta
)^{2}  \label{boson}
\end{equation}
where $\Theta $ and $\Pi =\nabla \Phi $ are canonical conjugate boson
fields; $u$ and $K$ are parameters giving respectively the velocity of the
harmonic wave and controlling anomalous exponents. We recall that the LL
hypothesis states that in one dimension any gapless hamiltonian $H$ (for
bosons, as well as for fermions and spins) will admit an effective
low-energy description in terms of the gaussian hamiltonian $H_{B}$ for a
suitable choice of the parameters $u$ and $K$ (these can be determined by
finite-size scaling methods or by comparison to exact solutions whenever
available). The mapping is completed through the relations $j=\frac{1}{\sqrt{%
\pi }}\nabla \Theta $ and $\delta \rho =\frac{1}{\sqrt{\pi }}\nabla \Phi $
where $j$ and $\delta \rho $ are respectively the particle current and a
density fluctuation about a mean value $\rho _{0}$, plus the definition of
the particle creation operators: $\Psi _{B}^{+}=\rho ^{1/2}\exp (i\sqrt{\pi }%
\Theta )$ for bosons, and $\Psi ^{+}=\Psi _{B}^{+}\quad (\exp (ik_{F}r+i%
\sqrt{\pi }\Phi )+\exp (-ik_{F}r-i\sqrt{\pi }\Phi ))$ for fermions. The
Fourier-transform of $H_{B}$ is: 
\begin{eqnarray}
H_{B} &=&\frac{u}{2}\sum_{q\neq 0}K^{-1}\Pi _{q}\Pi _{-q}+Kq^{2}\Theta
_{q}\Theta _{-q}  \nonumber \\
&&+\frac{\pi u}{2L}\left( \frac{\widehat{Q}^{2}}{K}+K\widehat{J}^{2}\right) 
\label{hamilton}
\end{eqnarray}
where $q$ is quantized as $q_{n}=2\pi n/L$, $\widehat{Q}=\widehat{N}-N_{0}$
counts particles from $N_{0}$ the ground state particle number and $\widehat{%
J}=\int j$ is a particle current. The ground state wavefunction of an
harmonic oscillator is a gaussian and is obtained for $N=N_{0}$ and $J=0$: $%
\Psi _{0}=\exp (-\frac{1}{2K}\sum_{n\neq 0}\frac{1}{\left| q_{n}\right| }\Pi
_{n}\Pi _{-n})$; returning to the original variables through $\Pi _{q}=\sqrt{%
\pi /L}\rho _{q}=\sqrt{\pi /L}\sum_{i}\exp (iqr_{i})$ , we easily find that $%
\Psi _{0}$ is nothing but a Jastrow-Laughlin wavefunction! 
\begin{equation}
{\psi }_{0}^{{}}(\{r_{1,..,}r_{N_{0}}\})=\prod_{i<j}\mid z_{ij}{\mid }^{1/K}
\label{bosonic}
\end{equation}
This is the correct form if we consider bosons; for fermions, we note that
the fermion creation operator is deduced from that of the boson by a
Jordan-Wigner phase factor multiplication which ensures antisymmetrization
(see Appendix); this means that reducing $H$ to $H_{B}$ for fermions is
achieved by first making a singular gauge transformation which converts the
particle statistics to a bosonic one. In the end, this transformation must
be undone and we obtain ${\psi }_{0}^{F}(\{r_{1,..,}r_{N}\})=\prod_{i<j}%
\left( z_{ij}\right) \mid z_{ij}{\mid }^{1/K-1}\exp ik_{F}\sum r_{i}+c.c.$.
This derivation of the ground state follows exactly the same lines as that
for the bosonic Landau-Ginzburg theory for the FQHE\cite{kivelson}. Neutral
excitations above the ground state (for which $Q=0$ and $J=0$) are Hermitte
polynomials: 
\begin{equation}
\left| n_{q_{1}},n_{q_{2}},...,n_{q_{p}}\right\rangle
=\prod_{s=1}^{p}H_{n_{q_{s}}}\left( \sum_{i}z_{i}^{n_{q_{s}}}/\sqrt{LK\left|
q_{s}\right| }\right) \left| \Psi _{0}\right\rangle   \label{hermitte}
\end{equation}
As usual for a harmonic oscillator they are obtained from ladder operators
which are here: $a_{q}=\frac{\Pi _{q}}{\sqrt{2K\mid q\mid }}-i\sqrt{\frac{%
K\mid q\mid }{2}}\Theta _{q}$. For instance, $\Psi _{n_{q}}\propto \rho
_{q}\Psi _{0}$. We note that this is just the excitation predicted by
single-mode approximation (SMA) theories, which were first considered by
Feynman in the context of superfluid $^{4}He$ \cite{sma}. These excitations
are the collective modes of the LL.

In this part we have derived the wavefunctions of the LL boson hamiltonian
for the ground state and its neutral excitations, which we have found to be
isomorphic to those considered for the FQHE. We will now show that this
parallel holds also for charged excitations.

{\bf {\centerline {Charged excitations in the LL.}}} The LL does not support
only density wave excitations : it is well known indeed from Bethe Ansatz
that there are spinons in the 1D Heisenberg model (spin $1/2$ excitations),
and additionally holons in the Hubbard model. The form of the LL ground
state invites us to write down the following variational quasihole
wavefunctions, in complete analogy with the FQHE: 
\begin{equation}
\Psi _{z_{0}}(x_{i})=\prod_{i=1}^{N}(z_{i}-z_{0})\prod_{i<j}\mid z_{ij}{\mid 
}^{1/K}
\end{equation}
As in two dimensions, from the plasma analogy we can show that $\Psi
_{z_{0}} $ carries a charge $K$: 
\begin{eqnarray}
\left| \Psi _{z_{0}}\right| ^{2} &=&\exp 1/K\int \int dydy^{\prime }\left[
\rho (y)+K\delta (x-y)\right]  \nonumber \\
&&\ln \left| \sin \frac{\pi }{L}(y-y^{\prime })\right| \left[ \rho
(y^{\prime })+K\delta (x-y^{\prime })\right]  \label{frac}
\end{eqnarray}
For $K=1/2$ (relevant to the Heisenberg model), that object has exactly the
spin expected for the spinon. We will now show that this guess is actually
the correct answer: a LL with parameter $K$ has excitations carrying charges
which are integer multiples of $K$.

Excitations can be classified according to the zero modes $\widehat{Q}$ and $%
\widehat{J}$ defined in eq.(\ref{hamilton}) which define (topological)
charge sectors $\{Q,J\}$: in each of these sectors states are obtained from
a lowest energy state by repeated application of the ladder operators $%
a_{q}^{+}$ which do not change the particle numbers. The identification of
charged excitations (for which $N\neq N_{0}$ and $J\neq 0$) is a standard
operation in conformal field theory (CFT), where they are obtained from so
called primary operators which generate the lowest energy eigenstates in a
given charge sector. For the gaussian theory, (a $c=1$ CFT), they are well
known to be vertex operators\cite{kadanoff,ginsparg}: 
\begin{equation}
V_{\alpha ,\beta }(x)=:\exp i\left( \sqrt{\pi }\beta \Phi (x)+\sqrt{\pi }%
\alpha \Theta (x)\right) :
\end{equation}
($::$ denotes normal ordering) which obey the commutation relations $\left[ 
\widehat{Q},V_{\alpha ,\beta }\right] =\alpha V_{\alpha ,\beta }$ and $%
\left[ \widehat{J},V_{\alpha ,\beta }\right] =\beta V_{\alpha ,\beta }$. We
interpret these excitations as longitudinal fluctuations of the superfluid
phase $\Theta $ and the $\exp (i\Phi )$ term as describing vortices. $%
J=\beta $ and $Q=\alpha $ are therefore topologically quantized and are
respectively the vortex circulation and the number of quasiparticles. They
are integers which we note now as $J=n$ and $Q=m$. Note that in a dual
language which exchanges vortex and particle variables, quasiparticle
excitations become (dual) vortices and the quantization of $Q$ is simply the
quantization of the circulation for the dual vortex.

In each topological sector the charged excitation with lowest energy is $%
V_{m,n}(z=0)$ with energy $\pi u(Q^{2}/K+KJ^{2})/2L$. For other values of
the parameter $z$, $V_{m,n}$ describes a coherent state living in the same
charge sector. As is often pointed out to demonstrate the non-Fermi liquid
nature of the LL, the electronic Green functions (i.e. for the $V_{m,n}$'s
which in the non-interacting problem represent $Q=m$ (bare) particles
carrying a current $J=n$) do not develop quasiparticle poles: it is believed
that only neutral excitations are fundamental due to a decay of (bare)
electrons. The fate of the electron is more subtle however.

The physical meaning of such operators is revealed when we go to first
quantization. Let us consider for instance $V_{1,0}$; using the identity
(for a review of the bosonization technique, see \cite{shankar}): $\Theta
(x)=\sum_{q}\theta _{q}e^{iqx}/\sqrt{L}=\sum_{q}sgn(q)\phi _{q}e^{iqx}/\sqrt{%
L}=-i/\pi \int_{0}^{L}dy\ln \left| \sin \frac{\pi }{L}(x-y)\right| \nabla
\Phi (y)$ we see that in first quantization: 
\begin{equation}
V_{1,0}=\exp \int_{0}^{L}\ln \left| \sin \frac{\pi }{L}(x-y)\right| \rho
(y)=\prod_{i=1}^{N_{0}}\left| z_{i}-z\right|
\end{equation}
Similarly, $V_{0,1}=\prod_{i=1}^{N_{0}}\left( z_{i}-z\right) /\left|
z_{i}-z\right| $ and $V_{1,1}(x)\Psi
_{0}(x_{i})=\prod_{i=1}^{N}(z_{i}-z)\prod_{i<j}\mid z_{ij}{\mid }^{1/K}$.
Therefore $V_{1,1}\Psi _{0}$ is precisely the quasihole wavefunction $\Psi
_{z_{0}}$ introduced above! This again is completely analogous to the FQHE.
We remark that $V_{0,1}$ precisely describes a phase singularity which
justifies its identification as a vortex. The elementary charge quantum was
determined in eq.(\ref{frac}) through plasma analogy and charges are
therefore integer multiples of $K$: $Q_{c}=Q$ $Ke$ where $Q=m$ is an integer
(this is the number of quasiparticles).

Quasi-electrons then are described by vertex operators with negative $m$: an
example is $V_{-1,-1}$, the effect of which amounts to dividing by a factor $%
\prod_{i=1}^{N}(z_{i}-z)$; we note that the description of anti-vortices in
the LL theory is restricted to a region excluding the core due to a
divergence when a particle comes close to $z$. This is a difficulty shared
by the effective gaussian Chern-Simons Landau-Ginzburg theory in 2D \cite
{kivelson} which does not affect however the correctness of the description
at long-distance.

The LL has a chiral symmetry. This has the additional consequence that the
hamiltonian can be separated into chiral components $H_{B}=H_{+}+H_{-}$: 
\begin{equation}
H_{\epsilon }=\frac{u}{2}\sum_{\varepsilon q>0}K^{-1}\Pi _{q}\Pi
_{-q}+Kq^{2}\Theta _{q}\Theta _{-q}+\frac{\pi u}{LK}\widehat{Q}_{\varepsilon
}^{2}  \label{chiral}
\end{equation}
where $\widehat{Q}_{\varepsilon }=\left( \widehat{Q}+\varepsilon K\widehat{J}%
\right) /2$ are chiral charges which count the number of particles within
each chiral sector: $\widehat{Q}=\widehat{Q}_{+}+\widehat{Q}_{-}$;
eigenvalues have the form: $Q_{\varepsilon }=(m+\varepsilon Kn)/2$ where $n$
and $m$ are arbitrary integers. Although there is a chiral separation of the
hamiltonian, it must be noted that the chiral charges of the allowed
excitations can not vary independently: they are constrained by the relation 
$Q_{\varepsilon }=(m+\varepsilon Kn)/2$. Since charged excitations mix
chiralities, one might be tempted to infer that chiral excitations are
confined (not separated): yet this is not correct. Rather one has a {\it %
pseudo-confinement}: although the true charged elementary excitations (the $%
V_{m,n}$) do appear as chiral composites, their chiral components are still
free, dynamically independent. The constraint $Q_{\varepsilon
}=(m+\varepsilon Kn)/2$ acts like a selection rule which has no bearing on
the chiral separation. We want to stress this point which will come out
again in the discussion of spin-charge separation. In the context of
spin-charge separation, that pseudo-confinement will appear with charged
excitations carrying both charge and spin although spin-charge separation is
indeed realized. The origin of these selection rules is the topological
quantization of $Q$ and $J$ the number of quasiparticles and their vorticity.

The fact that excitations come with both chiralities according to the above
constraint is very similar to the familiar topological constraint on spinons
in the Heisenberg chain which must come in pairs since they have a spin $1/2$%
\cite{fadeev}; they nevertheless are free particles. Since physical probes
can only involve integer numbers of particles, the only observable
variations of the spin and of the number of particles $\Delta S_{z}$ and $%
\Delta N$ must be integer-valued. This means that in a LL only charged
excitations carrying rational charges can be observed. Irrational charges $K$
are allowed only as parts of globally neutral complexes.

It is often stated that for $K<1$ (resp.$>1$), the LL describes repulsive
(resp.attractive) interactions. A simple examination of (\ref{bosonic})
allows a simple derivation of that result: when interactions are repulsive,
the ground state will develop higher-order zeros to keep particles further
apart. Now note that in the repulsive case the charge of the quasiparticle
will indeed be a fraction of the electron's unit charge; but in the
attractive case, our quasiparticle has a charge larger than that of the
electron: this reflects the attractive nature of the interactions though the
result is probably surprising.

We make additional remarks: (i) The boson hamiltonian displays a superfluid
rigidity for hard-core bosons. There is indeed a very striking parallel with
theories of $^{4}He$: we have a Jastrow ground state, vortex excitations
(the quasiparticles) and phonons (the neutral collective modes)\cite{sma}.
These hard-core bosons are obtained after a singular gauge transformation on
fermions: we may as in the FQHE understand the superfluidity as a hidden
off-diagonal long-range order\cite{girvin87}. (ii) The fractional
excitations $V_{m,n}$obey conventional exchange statistics which can be
shown to be: $\pi nm$\cite{mandelstam}, but anyons may however appear if we
consider generalizations of the LL to fields with conformal spin $S=(2n+1)/2$%
;(this will be discussed when we consider edge states). (iii) Still,
quasiparticles obey fractional exclusion statistics\cite{hal91} with
statistics parameter $K$; exclusion statistics is characterized by the fact
that one state can be occupied by at most one fermion or by any number of
bosons, but by $1/g$ particles obeying exclusion statistics with statistics $%
g$. For instance, for $V_{1,1}$ quasiparticles with charge $K=1/q$,
statistics is $1/q$ since $\Psi
_{0}(x_{1}..x_{N+1})=[V_{1,1}(x_{N+1})]^{q}\Psi _{0}(x_{1}..x_{N}).$ (iv) A
difficult problem in LL is the determination of the parameters $(u,K)$;
their numerical evaluation through the variational principle makes it a
simple matter: integrals with Jastrow-Laughlin functions can be done
straightforwardly with simple Metropolis algorithms. (v) Neutral collective
excitations which are bare particle-hole pairs may also be viewed as
Laughlin quasiparticle-quasihole pairs.

The structure of the excitation spectrum is summarized as follows: it is
given by the set of integers $\{N_{0},Q,J\}$ where for a given number of
particles $N_{0}$ we classify excitations in topological sectors $\{Q,J\}$.
In each sector there are topologically neutral excitations which are density
modulations. States in the sector $\{N_{0},Q,J\}$ carry a charge $N=N_{0}+KQ$%
; however $Q\neq 0$ is not allowed if $K$ is irrational since in the Fock
space $F=H(N=0)\oplus H(N=1)\oplus ...${\it \ }only states with a global
integer charge exist. Transitions to the sector $\{N_{0}+n,Q,J\}$ cannot
therefore be understood in terms of quasiparticle excitations from the
sector $\{N_{0},Q,J\}$ unless the integer $n$ is a multiple of $K$.

In summary we have shown in this part the novel result that Luttinger
liquids -among which we find important models such as the Heisenberg chain
or the Hubbard model- sustain very peculiar charged excitations which carry
indeed {\it anomalous} charges (i.e. non-integer in general), obey exclusion
statistics and which are just the 1D counterparts of the well-known Laughlin
quasiparticles. (A difference however is that charges in the FQHE are
rational numbers.) For instance the anisotropic Heisenberg model -eq.(\ref
{xxz})- has spin excitations carrying a spin $K=\frac{\pi }{2}/\arccos
(-\Delta ).$ As we vary the anisotropy $\Delta $ from $1$ to $0$, we observe
that the anomalous spin will vary from $1/2$ (the isotropic chain) to $1$ ($%
XY$ model) in a {\it continuous} manner. For ferromagnetic anisotropies ($%
\Delta <0$) the spin is larger than one (the interaction is indeed
attractive).

We conclude this part by a theorem: ''the wavefunctions of the exact
eigenstates of the LL boson hamiltonian are a Jastrow-Laughlin wavefunction
for the ground state, Laughlin quasiparticles for the charged excitations
and Bijl-Feynman phonons for the neutral excitations.'' {\bf {\centerline {%
Quantum numbers separation.}}} We now add internal quantum numbers to the
LL, and for definiteness focus on usual spins, considering the following
two-component wavefunctions: 
\begin{equation}
{\psi }_{0}(\{r_{i},\sigma _{i}\})=\prod_{i<j}\mid z_{ij}{\mid }^{g_{\sigma
_{i},\sigma _{j}}}
\end{equation}
where $\widehat{g}$ the charge matrix is a $2\times 2$ symmetric matrix ($%
\widehat{g}$ describes the charges of the classical plasma associated with $%
\Psi _{0}$). For fermions we antisymmetrize the wavefunction as follows: ${%
\psi }_{F}(\{r_{i},\sigma _{i}\})=\prod_{i<j}\mid z_{ij}{\mid }^{g_{\sigma
_{i},\sigma _{j}}-\delta _{\sigma _{i},\sigma _{j}}}(z_{ij})^{\delta
_{\sigma _{i},\sigma _{j}}}e^{i\frac{\pi }{2}sgn(_{\sigma _{i}-\sigma
_{j}})} $ (the exponential which ensures antisymmetrization between
different species is known as a Klein factor). We want to recover
spin-charge separation which is believed to be a characteristic of the
(''spinful'') LL. Quite often a problem difficult to handle in terms of
certain variables may become simple if one changes to dual variables: the
highly collective nature of the LL (anomalous charges can only exist on that
account) suggests to switch from individual particle coordinates to
collective ones -namely densities. Spin-charge separation is then indeed
readily apparent. We state that ${\psi }_{0}$ is the exact ground state of
the following hamiltonian for arbitrary velocities $u_{i}$: 
\begin{equation}
H(\widehat{g})=\sum_{i=1}^{2}\frac{u_{i}}{2}\int_{0}^{L}dx\
K_{i}^{-1}(\nabla \Phi _{i})^{2}+K_{i}(\nabla \Theta _{i})^{2}
\end{equation}
where the fields $\Phi _{i}$ are related to the densities $\rho _{\uparrow }$
and $\rho _{\downarrow }$ by $\rho _{i}=\nabla \Phi _{i}/\sqrt{\pi }$ and $%
\rho _{\sigma }=P_{\sigma i}\rho _{i}$ where $P$ is the unitary matrix which
puts $\widehat{g}$ to diagonal form with eigenvalues $K_{1}^{-1}$ and $%
K_{2}^{-1}$ , i.e. $P^{-1}\widehat{g}P=diag(1/K_{1},1/K_{2})$. We rewrite ${%
\psi }_{0}$ as: 
\begin{eqnarray}
&&\exp 1/2\int dydy^{\prime }\rho _{\sigma }(y)g_{\sigma \tau }\ln \left|
\sin \frac{\pi }{L}(y-y^{\prime })\right| \rho _{\tau }(y^{\prime }) 
\nonumber \\
&=&\exp \sum_{i}1/2K_{i}\int dydy^{\prime }\rho _{i}(y)\ln \left| \sin \frac{%
\pi }{L}(y-y^{\prime })\right| \rho _{i}(y^{\prime })
\end{eqnarray}
which proves the above statement and makes a separation into normal modes $%
\rho _{1}$ and $\rho _{2}$ manifest. We now see that spin-charge separation
can only occur if there is a $Z_{2}$ symmetry between up and down spins for $%
\widehat{g}$, i.e. if $\widehat{g}$ has the form $\left( 
\begin{array}{ll}
\lambda & \mu \\ 
\mu & \lambda
\end{array}
\right) $. Then $H(\widehat{g})$ is the usual spin-charge separated boson
hamiltonian with LL parameters $K_{\rho }=\frac{1}{\lambda +\mu }$ and $%
K_{\sigma }=\frac{1}{\lambda -\mu }$. Quite generally, if the gaussian
hamiltonian breaks this invariance, so will the ground state since it is
built from the normal modes of the hamiltonian; in that case, instead of
getting spin-charge separation one has a more general ''{\it quantum numbers
separation}''; normal modes will each carry parts of the quantum numbers of
the electron in a proportion {\it fixed} in time, and will factorize into
dynamically independent parts for all observables, which is a consequence of
the separability of the hamiltonian (at $T=0$ this can be seen equivalently
as separability of the ground state). That separation which is highly
non-trivial in terms of particles stems therefore from the rather trivial
statement that with $N$ internal degrees of freedom, there will be $N$
normal modes. (Separation into chiral components -see (\ref{chiral})- is
exactly similar.)

As an illustration of quantum numbers separation, we discuss the Hubbard
model in one dimension. For strong repulsion $U$, the Bethe Ansatz ground
state is known to factorize in spin and charge parts\cite{ogata}: $\Psi
\propto \prod_{i<j}(z_{ij})\Psi _{H}$ where $\Psi _{H}$ the spin part is
related to the Heisenberg chain Bethe Ansatz wavefunction, and the first
part (the charge part) is just a Slater determinant. Are we able to extract
the LL parameter $K_{\rho }$? Relating this form to the Jastrow-Laughlin
one, we can indeed read off $K_{\rho }=1/2$! In a magnetic field, from the
exact solution, it was argued\cite{frahm} that spin-charge separation is not
realized, and that the Hubbard model is a semi-direct product of two $c=1$
CFT's. A dressed charge matrix $Z$ describing renormalized charges of
excitations was introduced. That formalism can be related to the LL: it can
be shown that with the choice $\widehat{g}=M^{T}M$ where $M$ is related to $%
Z $ through $M=\left( 
\begin{array}{ll}
z_{cc}-z_{sc} & z_{sc} \\ 
z_{cs}-z_{ss} & z_{ss}
\end{array}
\right) $ where the $z_{ij}$ are $Z$ matrix elements, the anomalous
exponents predicted by Bethe Ansatz (plus CFT) are completely reproduced by
the two-component LL corresponding to $\widehat{g}$ \cite{moi}.This analysis
yields a $\widehat{g}$ matrix which breaks the $Z_{2}$ symmetry between up
and down spins, confirming the fact that spin-charge separation no longer
occurs when a magnetization sets in: but the model still maps onto a LL,
i.e. to $H(\widehat{g}=M^{T}M)$ which is the direct product of two $c=1$
CFT. We conclude that in a magnetic field the one dimensional Hubbard model
exhibits a quantum number separation (though no longer spin-charge
separation).

Charged excitations are again vertex operators: 
\begin{equation}
V_{m,n,m^{\prime },n^{\prime }}=:e^{i\sqrt{\pi }\left( n\Phi _{\uparrow
}+m\Theta _{\uparrow }+n^{\prime }\Phi _{\downarrow }+m^{\prime }\Theta
_{\downarrow }\right) }:
\end{equation}
where $J_{\sigma }=(n,n^{\prime })$ and $Q_{\sigma }=(m,m^{\prime })$ are
integer bare charges. Their exchange statistics is $\pi (mn+m^{\prime
}n^{\prime })$ and is therefore conventional. We rewrite eq.(12) as: 
\begin{equation}
{\psi }_{0}=\prod_{\uparrow }\mid w_{ij}{\mid }^{\lambda }\prod_{\downarrow
}\mid y_{ij}{\mid }^{\lambda ^{\prime }}\prod_{{}}\mid w_{i}-y_{j}{\mid }%
^{\mu }
\end{equation}
where $\{w_{i},y_{j}\}$ are the positions of spins up and down respectively.
(We have also set $g_{\uparrow \uparrow }=\lambda $, $g_{\downarrow
\downarrow }=\lambda ^{\prime }$ and $g_{\uparrow \downarrow }=\mu $.) For
example $V_{1,1,0,0}{\Psi }_{0}=\prod_{i}(w_{i}-z_{0}{)\Psi }_{0}$. We can
determine the charges $e_{\uparrow }$ and $e_{\downarrow }$ of the vertex
operators by plasma analogy again: $\left| V_{Q_{\sigma },J_{\sigma }}\Psi
_{0}\right| ^{2}=e^{-U}$ where: 
\begin{eqnarray}
U &=&\int \int \left[ \rho _{\sigma }(y)+g_{\sigma \tau }^{-1}Q_{\tau
}\delta (x_{0}-y)\right] g_{\sigma \sigma ^{\prime }}  \nonumber \\
&&\ln \left| \sin \frac{\pi }{L}(y-y^{\prime })\right| \left[ \rho _{\sigma
^{\prime }}(y^{\prime })+g_{\sigma ^{\prime }\tau }^{-1}Q_{\tau }\delta
(x_{0}-y^{\prime })\right]
\end{eqnarray}

The charges are therefore: $e_{\sigma }=g_{\sigma \tau }^{-1}Q_{\tau }$. For
instance for $V_{1,1,0,0}$, $Q_{\sigma }=(1,0)$ and the charges are: $%
(e_{\uparrow },e_{\downarrow })=\left( \frac{\lambda ^{\prime }}{\lambda
\lambda ^{\prime }-\mu ^{2}},\frac{-\mu }{\lambda \lambda ^{\prime }-\mu ^{2}%
}\right) $. These charges can also be determined algebraically from the zero
modes of the normal densities $\rho _{i}$: in the normal basis, they are
again $e_{i}=K_{i}Q_{i}$ (in the spinless case we identified the anomalous
charge as $KQ$); going back to the spin basis, $e_{\sigma }=P_{\sigma
i}K_{i}Q_{i}$. Since $Q_{i}=P_{\tau i}Q_{\tau }$ (because $\rho
_{i}=P_{\sigma i}\rho _{\sigma }$), we obtain again $e_{\sigma }=P_{\sigma
i}K_{i}P_{\tau i}Q_{\tau }=g_{\sigma \tau }^{-1}Q_{\tau }$. The vertex
operators $V_{Q_{\sigma },J_{\sigma }}$ therefore carry a charge $%
q=e_{\uparrow }+e_{\downarrow }=\sum_{\sigma \tau }g_{\sigma \tau
}^{-1}Q_{\tau }$ and a spin $S_{z}=\frac{1}{2}(e_{\uparrow }-e_{\downarrow
})=\sum_{\sigma \tau }\sigma g_{\sigma \tau }^{-1}Q_{\tau }/2$. As discussed
in the context of chiral separation, although charged excitations carry both
anomalous charges $K_{1}Q_{1}$ and $K_{2}Q_{2}$, there is still quantum
numbers separation: this again is not a true confinement, but rather a
pseudo-confinement.

We specialize the discussion to the case of spin-charge separation: $\lambda
=\lambda ^{\prime }$. We rewrite $V_{1,1,0,0}=\exp \int \rho _{\uparrow
}(x)\ln (z-z_{0})dx$ as: 
\begin{equation}
e^{\frac{1}{2}\int \rho _{c}(x)\ln (z-z_{0})dx}e^{\frac{1}{2}\int \rho
_{s}(x)\ln (z-z_{0})dx}  \label{factor}
\end{equation}
The two factors are dynamically independent due to spin-charge separation of
the hamiltonian and correspond respectively to the holon and to the spinon.
We can also express the holon as $h(z_{0})=V_{1/2,1/2,1/2,1/2}=%
\prod_{i}(w_{i}-z_{0}{)}^{1/2}\prod_{i}(y_{i}-z_{0}{)}^{1/2}$ and the spinon
as $s(z_{0})=V_{1/2,1/2,-1/2,-1/2}=\prod_{i}(w_{i}-z_{0}{)}%
^{1/2}/\prod_{i}(y_{i}-z_{0}{)}^{1/2}$. They carry respectively charge and
spin $(q=K_{\rho },s=0)$ and $(q=0,s=K_{\sigma }/2)$. (In the $SU(2)$
symmetric case, $K_{\sigma }=1$ and the spinon has spin $1/2$ as expected.)
More generally $V_{m,n,m^{\prime },n^{\prime }}$ carries charge $%
q=(m+m^{\prime })K_{\rho }$ and spin $s=(m-m^{\prime })K_{\sigma }/2.$ Note
that the holon and the spinon are semions with statistics $\pi /2$. Due to
pseudo-confinement however, the total number of holons and spinons is always
even (as can be inferred from the expression for the charge and the spin of $%
V_{m,n,m^{\prime },n^{\prime }}$) and therefore neither parity $P$ nor time
reversal $T$ are ever broken: we come to the surprising conclusion that we
can have $P$ and $T$ breaking excitations although a (global) $P$ and $T$
breaking will never be observed. We stress that these pairs of holons and/or
spinons are {\it not} bound. Such is not the case because they are
dynamically independent: they depend on charge and spin densities which have
independent dynamics due to the separation of the hamiltonian. The
excitations appear as composites only because the total topological charges $%
Q$ and $J$ must be integers. The wavefunctions of the allowed charged
excitations (and of the ground state) will factorize into independent charge
and spin parts corresponding to the holons and spinons.

In summary we have found that spin-charge separation is a very natural
property of Jastrow-Laughlin wavefunctions; we have generalized the usual
spin-charge separation to a ''quantum numbers separation'' into independent
normal modes (which mix charge and spin in a proportion {\it fixed} in
time). As in the spinless case we have argued again that there is a
pseudo-confinement for charged excitations.

{\bf {\centerline {Jastrow-Luttinger Fractional Liquids.}}} We now
systematicaly compare the LL and the FQHE, which leads us to formulate the
concept of a Jastrow-Luttinger fractional liquid. (i) First of all, the
ground states $\Psi _{0}$ have the same functional form and describe
featureless liquids with a uniform density. (ii) Branches of neutral
collective excitations correspond to $\rho _{k}\Psi _{0}$\cite{plat}: they
are density fluctuations above the liquid surface, and can be viewed
therefore as bare particle-hole pairs or as anomalous charge
quasiparticle-quasihole pairs. (iii) Charged excitations are solitons
corresponding to bumps or holes at the surface of the liquid, i.e. Laughlin
quasiparticles. (iv) In both cases, {\it gaussian }effective Landau-Ginzburg
theories can be written: in the FQHE at filling $\nu =1/(2n+1)$ starting
from a microscopic hamiltonian, after a Chern-Simons transformation one can
derive an effective Landau-Ginzburg theory; after integration of the
Chern-Simons gauge field one obtains the following hamiltonian\cite{kivelson}%
: 
\begin{equation}
H=\frac{\rho _{0}}{2m}\sum_{q}\left( \frac{2\pi }{\nu q}\right) ^{2}\Pi
_{q}\Pi _{-q}+q^{2}\Theta _{q}\Theta _{-q}
\end{equation}
where $\Pi _{q}\propto \delta \rho _{q}$ (the density fluctuations) is the
canonical conjugate of $\Theta $,(compare with (\ref{boson})) and whose
exact ground state is Laughlin wavefunction at filling $\nu =1/(2n+1)$. (v)
For fermions, in both instances, reduction to the gaussian theory is done
after a change in statistics (CS transformation in 2D, Jordan-Wigner
transformation in one dimension): we then end up with Landau-Ginzburg
theories for hard-core bosons (the composite bosons of the FQHE in two
dimensions). (vi) Both hamiltonians exhibit a superfluid rigidity , with a
quasi Off-Diagonal Long-Range Order (ODLRO) for the Laughlin bosons (as
stressed in two dimensions in\cite{girvin87}): it is easily seen from the
boson hamiltonians that $\left\langle \Psi _{B}(0)\Psi
_{B}^{+}(x)\right\rangle \sim 1/|x|^{1/2K}$ (with $K$ replaced by $\nu $ in
2D). The LL is a critical theory with algebraic decay for all order
parameters. Note however that for attractive interactions ($K>1$)
superfluidity of Laughlin bosons is always the dominant order parameter,
while for repulsive interactions ($K<1$), it is superfluidity for the dual
variable (vortices) which then has the slower decay ($\left\langle e^{i\sqrt{%
\pi }\phi (0)}e^{i\sqrt{\pi }\phi (x)}\right\rangle \sim 1/|x|^{K/2})$.
Therefore either one or the other of these two superfluidities is always the
dominant order. (vii) Charged excitations obey fractional exclusion
statistics\cite{hal91}. (viii) When we consider multicomponent systems, in
both cases again, there will be a quantum-numbers separation.

We then define Jastrow-Luttinger Fractional Liquids to be superfluids with
Jastrow-Laughlin ground states, Laughlin quasi-particles carrying anomalous
charges, Bijl-Feynman collective modes and displaying a quantum number
separation. A striking point when we consider fermions is that the ODLRO
characteristic of superfluidity is {\it hidden}: it appears after a singular
gauge transformation converting the statistics of fermions into a bosonic
one (Chern-Simons or Jordan-Wigner transmutation). This peculiar hidden
order was first emphasized by Girvin and MacDonald in \cite{girvin87} for
the FQHE and explains the strong analogy with the physics of $^{4}He$
(Jastrow ground states, vortices, phonons, rotons): as shown above that
property is very remarkably shared by the LL which we re-interpret as a
hard-core boson superfluid.

We now prove spin-charge separation for the FQHE; as early as 1983, Halperin
introduced the following two-component wave functions\cite{halperin}: 
\begin{equation}
{\phi }_{m,m^{\prime },n}=\prod_{i<j}(z_{ij}{)}^{g_{\sigma _{i},\sigma
_{j}}}\prod_{i}\exp -\frac{\left| z_{i}\right| ^{2}}{4}
\end{equation}
with $\widehat{g}=\left( 
\begin{array}{ll}
m & n \\ 
n & m^{\prime }
\end{array}
\right) $ in order to describe non-fully polarized states in the QHE at
fillings $\nu =(m+m^{\prime }-2n)/(mm^{\prime }-n^{2})$. Such wavefunctions
are Laughlin's state natural extensions to multicomponent systems and have
been widely used to describe spin effects and multilayer systems. For $%
m=m^{\prime }=n+1$, we have singlet states \cite{component}. As in the
single component case these wavefunctions can be derived microscopically
from bosonic Chern-Simons Landau-Ginzburg (CSLG) theories; for instance one
may check that , $\phi _{B}=\left| {\phi }_{m,m,n}\right| $ is the exact
ground state of: 
\begin{eqnarray}
H &=&\frac{\rho _{0}}{2m}\sum_{q,\sigma }\left( \frac{2\pi }{q}\right)
^{2}(m+n)^{2}\Pi _{q,\sigma }\Pi _{-q,\sigma }  \nonumber \\
&&+4mn\Pi _{q,\uparrow }\Pi _{-q,\downarrow }+q^{2}\Theta _{q,\sigma }\Theta
_{-q,\sigma }  \label{fqhe}
\end{eqnarray}
at filling $\nu =2/(m+n)$ for $m\neq n$ \cite{ezawa}. (Following eq.(12) a
similar generalization to the case $m\neq m\prime $ can also be written.)
From the previous analogous discussion of the LL, it should be clear hovewer
that ${\phi }_{m,m^{\prime },n}$ and $H$ for $m=m^{\prime }$ are both
spin-charge separated! More generally, again there will be a quantum numbers
separation, with a separation for the hamiltonian, the ground state and all
observables: the quantum numbers of the electron have vanished from the
excitation spectrum.

A possible worry in the FQHE is the requirement of lowest Landau level
projection: it is well known that CSLG theories do not adequately describe
the neutral modes since they do not work with projected density operators.
Still, using a full first quantized approach to multicomponent FQHE systems
removes the problem \cite{sondhi,component}: the ground state is taken as ${%
\phi }_{m,m^{\prime },n}$, charged excitations are $\prod_{i}(w_{i}-z_{0}{%
)\phi }_{m,m^{\prime },n}$ or $\prod_{i}(y_{i}-z_{0}{)\phi }_{m,m^{\prime
},n}$ (with notations similar to the 1D case), and neutral excitations are
found which for $m=m^{\prime }$ are precisely $P\rho _{c}{\phi }%
_{m,m^{\prime },n}$ and $P\rho _{s}{\phi }_{m,m^{\prime },n}$, i.e. charge
and spin modes again ($P$ is the lowest Landau level projector). This shows
that lowest Landau level projection is immaterial to the issue of
spin-charge separation.

We note that Laughlin quasiparticles are the only topologically allowed
charged excitations: in analogy to the LL we propose to define $Q$ and $J$
as the number of quasiparticles and the circulation of a vortex (in units of 
$\phi _{0}$); integer valuedness of these topological charges is only
realized for Laughlin quasiparticles. More precisely we write Laughlin
quasiparticles as 2D vertex operators: $V_{Q,J}^{2D}(z_{0})=\prod_{i}\left[
\left( z_{i}-z_{0}\right) /\left| z_{i}-z_{0}\right| \right] ^{J}\left|
z_{i}-z_{0}\right| ^{Q}$ (a difference with the 1D case is that $Q=J$ to
implement analyticity). Then all the considerations made above for the
charged excitations of a LL can be adapted to the FQHE with the formal
replacement: $z=\exp i\frac{2\pi }{L}x\longrightarrow z=x+iy$. For instance $%
(e_{\uparrow },e_{\downarrow })=\left( \frac{m^{\prime }}{mm^{\prime }-n^{2}}%
,\frac{-n}{mm^{\prime }-n^{2}}\right) $ for $\prod_{i}(w_{i}-z_{0}{)\phi }%
_{m,m^{\prime },n}$ which is associated with the 2D vertex operator $%
V_{1100}^{2D}=\prod_{i}(w_{i}-z_{0}{)}$; more generally, the charges carried
are recovered as $e_{\sigma }=g_{\sigma \tau }^{-1}Q_{\tau }$ for $%
V_{Q_{\sigma },J_{\sigma }}^{2D}$. Again it is clear that the quasiparticle
wavefunctions can be written as products of independent parts as in eq.(\ref
{factor}) (charge and spin parts if $m=m^{\prime }$) and with a
pseudo-confinement. In other words, for $m=m^{\prime }$ (but $m\neq n$ which
is a special case due to spin degeneracy -see below) we find that holon and
spinon excitations exist although the system is {\it two} dimensional; a
difference with the LL however is that both excitations are gapped.
Explicitly, the holon and spinon wavefunctions are respectively: $h(z_{0}){%
\phi }_{m,m,n}=\exp \frac{1}{2}\int \rho _{c}(x)\ln (z-z_{0})dx$ ${\phi }%
_{m,m,n}$ and $s(z_{0}){\phi }_{m,m,n}=\exp \frac{1}{2}\int \rho _{s}(x)\ln
(z-z_{0})dx$ ${\phi }_{m,m,n}$ (these expressions show explicitly the
independence of holons and spinons if the system is indeed described by a
boson hamiltonian with a separation in charge and spin densities -as in (\ref
{fqhe})). The role of the LL parameter $K$ is played by the filling factor $%
\nu $; for instance in the spin-charge separated case $K_{\rho
}\Longleftrightarrow (\nu _{\uparrow }+\nu _{\downarrow })/2$ and $K_{\sigma
}\Longleftrightarrow (\nu _{\uparrow }-\nu _{\downarrow })/2$. Charge and
spin of the holon and spinon are as in 1D after such formal replacements in
the previous formulas. As in the 1D case holons and spinons are free
although they appear together in the Laughlin quasi-hole wavefunction $%
\prod_{i}(w_{i}-z_{0}{)\phi }_{m,m,n}=h(z_{0})s(z_{0}){\phi }_{m,m,n}$ due
to the constraint on the total topological charges $Q_{\sigma }$ and $%
J_{\sigma }$ which must be integer valued.

We observe also that the charge and spin parts of the wavefunctions of the
ground state and of the charged excitations (e.g. $h(z_{0})$) are not
analytical functions, in violation of lowest Landau level projection: this
is no cause for concern because of pseudo-confinement. Indeed the only
requirement is that the full wavefunction be analytical; by virtue of
pseudo-confinement the spin and charge parts (spinons and holons) are never
observed separately although they are still independent because of
spin-charge separation. In short, the description of the multicomponent FQHE
-either in first quantization or in a CSLG approach- supports quantum
numbers separation in 2D in full parallel to the LL.

We expect that analysis to break down if there is a degeneracy of the ground
state because of the possibility of novel excitations (textures). And indeed
for ${\phi }_{m,m,m}$ (i.e. $m=m^{\prime }=n)$ which describes a QH
ferromagnet\cite{component} we obtain a singular (non invertible) $\widehat{g%
}$ matrix. But the filling fractions $\nu _{\sigma }$, the charges of the
Laughlin quasiparticles $e_{\sigma }$ and therefore their spin $S_{z}$ are
determined by $\widehat{g}^{-1}$ and therefore are ill-defined: this just
reflects the spin degeneracy. That degeneracy means that the above
description for the excitations of the system is insufficient since we can
consider textures interpolating between two ground states: this is precisely
the skyrmion excitation introduced for QH ferromagnets. Since this lies out
of the above theoretical framework there is no reason to demand spin-charge
separation.

Turning to experiments, we look for fillings suitable to test
quantum-numbers separation; it is known that for many fillings the quantum
Hall state is not fully polarized \cite{eisenstein}: this is explained by
the small effective masses and Zeeman couplings observed in real samples,
which can increase the ratio of the cyclotron to the Zeeman gap up to a
factor of $50$ (as in $GaAs$). Most promising for the observation of
spin-charge separation are fillings such as $\nu =8/5$ or $2/3$ for which
experimental evidence points to non-polarized ground states (although they
are not inconsistent with partially polarized ones) as seen in tilted-field
experiments \cite{eisenstein} (where transitions to polarized states are
observed as the in-plane field increases). Further support for a
non-polarized ground state at $\nu =8/5$ comes from numerical results at the
particle-hole conjugate filling $\nu =2/5$ (identical ground states are
expected after particle-hole conjugation): the exact ground state was shown
to be unpolarized for small values of the Zeeman coupling $g$, while
Halperin's unpolarized $\phi _{3,3,2}$ was also shown to have a very good
variational energy (better than the fully polarized hierarchical state),
although unfortunately its overlap with the true ground state was not
examined \cite{chakra}.

How could spin-charge separation be experimentally tested? Given that
separation in the bulk automatically entails separation at the edges,
experiments at the edge would confirm quantum-numbers separation for edges
and provide strong support in favor of its existence in two dimensions
(reflecting our claim that ground state wavefunctions have identical
functional forms both at the edges and in the bulk). Various tests were
proposed for the specific case of the LL: probing the spectral density which
has a two-peak structure due to spin-charge separation\cite{voit}, spin
injection which probes spin transport\cite{si}. In a different context, for $%
\nu =2/3$ hierarchical spin polarized edges, where some theories predict a
neutral collective mode besides a charged one, Kane and Fisher suggested
time domain experiments, in which an electron is injected through a tunnel
junction at the edge of a FQHE disk\cite{kane}. This proposal evidently
applies for the detection of charge and (real) spin modes since the latter
is neutral. Upon injection of the electron, the collective charge and spin
modes are excited but since they propagate at different speeds a detector
(another tunnel junction) would see two pulses. If we now replace the
junction by a capacitor, one will detect only the charge mode thereby
proving the existence of a neutral mode and of spin-charge separation, which
is relevant for such fillings as $\nu =8/5$. {\bf {\centerline {Edge states
in the FQHE.}}} We now turn to a discussion of edge states in the FQHE,
where the strong unity between the LL and the FQH liquid as
Jastrow-Luttinger Fractional liquids is further evidenced. At the edge of QH
samples, gapless states must exist since the Fermi level crosses the
confining potential\cite{halperin2}. On general grounds (CS theories and
hydrodynamical theories) Wen argued convincingly that these edge states are
described by a chiral LL\cite{wen}. The relation to Laughlin microscopic
approach remained obscure hovewer. Let us try to see how this may come
about: we will make use of two operations, one dimensional restriction of
the bulk wavefunction and analyticity for lowest Landau level states. (The
first operation can be given a precise meaning as is well known under the
namesake of one dimensional reduction in the lowest Landau level, or from a
direct reduction of the microscopic hamiltonian. See \cite{capelli} and
references therein. For our purposes, the following physical argument will
be enough: since bulk excitations lie above a gap, while edge states are
gapless, the latter decouple dynamically from the bulk for sufficiently
low-energy processes: bulk and edge can be thought of as living in two
decoupled Hilbert spaces, which gives meaning to a restriction of bulk
wavefunctions to the edges.)

From the above discussion of the LL, the relation between Laughlin
microscopic variational approach and the edge theories should be intuitively
clear: heuristically, in the bulk we have a Laughlin wavefunction and
therefore at the edge we also have a Jastrow-Laughlin wavefunction, which is
precisely the variational wavefunction associated with the LL; consequently
we have a LL at the edges, chirality stemming from the magnetic field.
However it must be noted that the bulk wavefunction is analytic while the
fermionic Luttinger wavefunction we gave above was not. This is easily
remedied by generalizing our previous treatment of the LL as follows: an
implicit assumption for the fermion fields $\Psi =\exp (i\sqrt{\pi }\Theta
)\quad (\exp (ik_{F}r+i\sqrt{\pi }\Phi )+\exp (-ik_{F}r-i\sqrt{\pi }\Phi ))$
was made, namely that they had conformal spin $1/2$ (conformal spin for
operators $V_{\alpha \beta }$ is $S=\pi \alpha \beta /2$; after a Wick
rotation, it describes the usual spin in Euclidean space-time); we may
consider however higher spins such as $S=(2n+1)/2$. This can be done by
''flux'' attachment (iteration of Jordan-Wigner transformations): $\Psi
^{\prime }=\exp (i\sqrt{\pi }2n\Phi )\Psi $, or by a modification of the
relations beween particle currents and the fields: $\delta \rho =\nabla \Phi
/\sqrt{\pi (2n+1)}$ and $j=-\nabla \Theta /\sqrt{\pi (2n+1)}$ (in technical
CFT jargon, we have modified the $U(1)$ Kac-Moody relations which define the
LL) with the new definitions $\Psi =\exp (i\sqrt{\pi (2n+1)}\Theta )\quad
(\exp (ik_{F}r+i\sqrt{\pi (2n+1)}\Phi )+h.c.)$. The first operation simply
leads to the following fermionic ground state: ${\psi }_{0}^{F}(%
\{r_{1,..,}r_{N}\})=\prod_{i<j}(z_{ij}/|z_{ij}|)^{2n+1}\mid z_{ij}{\mid }%
^{1/K}\exp ik_{F}\sum r_{i}+c.c.$ which under a $2\pi $ rotation of $z$
coordinates yields indeed a phase $4\pi S$ with $S=(2n+1)/2$; we will follow
the second prescription which is relevant for edge states: by retracing the
steps taken to obtain equ.(6) the second operation is easily shown to lead
to ${\psi }_{0}^{F}(\{r_{1,..,}r_{N}\})=\prod_{i<j}(z_{ij}/|z_{ij}|)^{2n+1}%
\mid z_{ij}{\mid }^{2n+1/K}\exp ik_{F}\sum r_{i}+c.c.$

If we restrict now the 2D $\nu =1/(2n+1)$ Laughlin bulk ground state to 1D,
this corresponds to the ground state of a LL with modified Kac-Moody
relations if we choose $K=1$ and if we drop the anti-analytical part of the
wavefunction, i.e. if the LL is chiral. Indeed implementing the analyticity
constraint on the excitations of the full non-chiral LL means that we have
to restrict to $q>0$ in (\ref{hermitte}) and to $\alpha =\beta $ for the $%
V_{\alpha ,\beta }$ vertex operators. This then means that these excitations
are chiral since $q>0$ while $\alpha =\beta $ implies that one of the chiral
charges is always zero. We observe therefore that lowest-Landau level
analyticity requirement and one-dimensional reduction alone imposed on
wavefunctions allow one to recover a chiral LL. We also note that the
operator whose action is $\prod_{i}(z_{i}-z)$ is $V_{\alpha ,\beta }$ with $%
\alpha =\beta =1/\sqrt{(2n+1)}$. It has therefore anyonic statistics $\pi
/(2n+1)$ as expected (that operator is allowed if we are on an annulus).
(The operation of flux attachment we first described above as an alternative
to modifying Kac-Moody relations can e shown to lead to a gapless
fractionally charged excitation, which is not allowed for a single FQH edge
on a disk geometry since Laughlin quasiparticles are gapped excitations;
this is the reason why we have considered the second alternative.) {\bf {%
\centerline {Spin liquids.}}} We now want to point out an interesting result
for quantum magnetism: if we compute the spin correlators for Halperin's $%
\phi _{m+1,m+1,m}$ which is a $SU(2)$ spin singlet, we find from the
gaussian hamiltonian - eq.(21) - that the longitudinal (and due to
rotational invariance, transverse as well) spin-spin correlators behave as $%
\left\langle S_{q}^{z}S_{-q}^{z}\right\rangle \sim q^{2}$ showing that
spin-spin correlations are suppressed. We have a local singlet state for
which spins are completely screened (this can be seen by exploiting
spin-charge separation: the spin part was studied in \cite{semion} and
represents in the plasma analogy a two component neutral Coulomb gas in the
disordered phase of the Kosterlitz-Thouless transition). We have therefore a
(spin-charge separated) spin liquid! This supports Kalmeyer-Laughlin
proposal of using Laughlin-like wavefunctions to describe spin-liquids\cite
{kalmeyer}. We note that we have thus two classes of spin liquids since in
one dimension gapless spin liquids are also described by Jastrow-Laughlin
wavefunctions. We stress the non-triviality of these statements: apart from
exact Bethe Ansatz states, very few fundamental magnetic states are known
for quantum antiferromagnetism. Indeed the N\'{e}el state (and its Ising
variant) stand out almost alone. We now have in low dimensions to add to
that list the Laughlin state. We note also that these spin-liquids are
Mott-Hubbard insulators: in the bulk, the $\phi _{m+1,m+1,m}$ which is a
state induced by the repulsive Coulomb interactions is clearly insulating
(there is a gap to charged excitations and a zero conductivity), with the
additional peculiarity that it is a spin charge separated Mott-Hubbard
insulator. \vskip 0.3cm We summarize the novel results obtained in this
paper: we have shown that the Luttinger liquid can be understood in terms of
Jastrow-Laughlin states, Laughlin quasiparticles, and Bijl-Feynman phonons.
As in the FQHE anomalous charges are sustained by the LL. We have introduced
the concept of pseudo-confinement for condensed-matter systems and
generalized spin-charge separation to two dimensions for the FQHE (for
situations without spin-degeneracy). We have characterized the LL and the
FQH fluid as Jastrow-Luttinger Fractional Liquids, i.e. superfluids with a
hidden off-diagonal-long-range-order for Laughlin bosons.

In low dimensions interactions and quantum fluctuations conspire to
stabilize novel quantum liquids : in that respect Jastrow-Luttinger liquids
which comprise the Luttinger liquid and the FQH fluid are paradigmatic of
strongly correlated systems. Such systems describe liquids which develop
Jastrow-Laughlin correlations to minimize interactions leading to a novel
kind of quantum coherence precisely described in a mean-field like manner by
Jastrow-Luttinger states. This special coherence means that the spectrum of
excitations is entirely collective: usual quasiparticles have vanished
(orthogonality catastrophe) through a quantum numbers separation and through
fractional charged excitations describing density bumps and holes above the
ground state sea. Fractionalization is complete. Such liquids are
astonishingly diverse: there are strange non-Fermi liquids metals yet with a
Fermi surface, Mott-Hubbard insulators or spin liquids. We end up with a
remark on high-temperature superconductivity: while the ideas of a
spin-liquid and of spin-charge separation proposed by Anderson\cite{anderson}
are quite controversial issues, we believe that this work contributes to
validate these concepts theoretically (and in the near future hopefully
experimentally) in two dimensions, albeit in the different context of
Jastrow-Luttinger Fractional Liquids. Whether the paradigm set by
Jastrow-Luttinger fractional liquids will eventually prove useful for the
understanding of high-temperature superconductivity remains an open
question. The authors thank B.Jancovici, D.Bazzali for interesting
discussions, as well as the Orsay theory group for stimulating interactions.

{\bf {\centerline {Appendix: statistical transmutations.}}}We show that the
Jordan-Wigner transformation is just the 1D restriction of the 2D
Chern-Simons transformation. The CS operator reads: 
\[
U(z)=\exp i\int dz^{\prime }\widehat{\rho }(z^{\prime })\arg (z-z^{\prime }) 
\]

where $\arg (z)$ is the argument of the complex variable $z$. Applied on a
fermionic operator, it transforms it into a hard-core boson. If $z$ and $%
z^{\prime }$ are constrained to lie on the real line, then $\arg
(z-z^{\prime })=\pi \theta (x-x^{\prime })$ where $\theta $ is the step
function. $U(z)$ becomes: 
\begin{eqnarray*}
U(x) &=&\exp i\int dx^{\prime }\widehat{\rho }(x^{\prime })\pi \theta
(x-x^{\prime }) \\
&=&\exp i\pi \int^{x}dx^{\prime }\widehat{\rho }(x^{\prime })
\end{eqnarray*}

which is just the continuum version of the usual Jordan-Wigner
transformation: 
\[
U_{n}=\exp i\pi \sum_{j<n}c_{j}^{+}c_{j} 
\]

In the bosonization formalism, $U(x)=\exp i(k_{F}x+\sqrt{\pi }\phi (x))$
since $\widehat{\rho }(x)=k_{F}/\pi +\nabla \phi /\sqrt{\pi }$. In first
quantization the action of both $U(z)$ and its 1D restriction $U(x)$ are
simply to multiply the wavefunction by the phase factor $\Pi
_{i}(z_{i}-z)/|z_{i}-z|$. We note also that spins $1/2$ are hard-core
bosons; therefore they can be fermionized in 1D and 2D by using
Jordan-Wigner or Chern-Simons transformations.

\end{document}